\documentclass[twocolumn,,secnumarabic,superscriptaddress,notitlepage, 11pt,longbibliography,reprint]{revtex4-1}

\usepackage{ulem}
\usepackage{mathtools}
\usepackage[colorinlistoftodos,prependcaption,textsize=tiny]{todonotes}

\begin{document}
\title{Numerical analysis of beam self-cleaning in multimode fiber amplifiers}
\author{Mesay Addisu Jima}
\affiliation{Dipartimento di Ingegneria dell'Informazione, Universit\`{a} di Brescia, via Branze 38, 25123 Brescia, Italy}
\affiliation{Universit\'e de Limoges, XLIM, UMR CNRS 7252, 123 Av. A. Thomas, 87060 Limoges, France}

\author{Alessandro Tonello}
\affiliation{Universit\'e de Limoges, XLIM, UMR CNRS 7252, 123 Av. A. Thomas, 87060 Limoges, France}

\author{Alioune Niang}
\altaffiliation[Present address: ]{Computer Science and Electrical Engineering Department, University of Maryland, Baltimore, MD 21250, USA}
\affiliation{Dipartimento di Ingegneria dell'Informazione, Elettronica e Telecomunicazioni, Sapienza Universit\`{a} 
di Roma, via Eudossiana 18, 00184 Rome, Italy}

\author{Tigran Mansuryan}
\affiliation{Universit\'e de Limoges, XLIM, UMR CNRS 7252, 123 Av. A. Thomas, 87060 Limoges, France}

\author{Katarzyna Krupa}
\affiliation{Institute of Physical Chemistry Polish Academy of Sciences, ul. Kasprzaka 44/52, 01-224 Warsaw, Poland}

\author{Daniele Modotto}
\email[Corresponding author: ]{daniele.modotto@unibs.it}
\affiliation{Dipartimento di Ingegneria dell'Informazione, Universit\`{a} di Brescia, via Branze 38, 25123 Brescia, Italy}

\author{Annamaria Cucinotta}
\affiliation{Dipartimento di Ingegneria e Architettura, Universit\`{a} di Parma, 
Parco Area delle Scienze 181/A, I-43124 Parma, Italy}

\author{Vincent Couderc}
\affiliation{Universit\'e de Limoges, XLIM, UMR CNRS 7252, 123 Av. A. Thomas, 87060 Limoges, France}

\author{Stefan Wabnitz}
\affiliation{Dipartimento di Ingegneria dell'Informazione, Elettronica e Telecomunicazioni, Sapienza Universit\`{a} 
di Roma, via Eudossiana 18, 00184 Rome, Italy}
\affiliation{CNR-INO, Istituto Nazionale di Ottica, Via Campi Flegrei 34, 80078 Pozzuoli (NA), Italy}

\begin{abstract}
Recent experimental results have reported the observation of beam self-cleaning or, more generally, 
nonlinear beam reshaping in active multimode fibers. In this work we present a numerical analysis 
of these processes, by considering the ideal case of a diode-pumped signal amplifier made of a graded-index multimode fiber with uniform Yb doping. Simulations confirm that beam cleaning of the signal may take place 
even in amplifying fibers, that is the absence of beam energy conservation. Moreover, 
we show how the local signal intensity maxima, which are periodically generated by the self-imaging process, may influence the population inversion of the doping atoms, and locally saturate the amplifier gain. 
\end{abstract}

\maketitle

\section{Introduction}

Light propagation in active waveguides may hold surprises or counter-intuitive situations. An example is provided by the
robust singlemode behaviour which can be obtained, in the presence of sufficiently large gain, even in step-index fibers with a refractive index dip in the core, or in anti-guides \cite{bib:Siegman:07}. Doped fibers provide the most technologically successful, and suitable platform for optical amplifiers. Accurate models to describe short pulse propagation in active fibers, including the contribution to dispersion introduced by the active ions, have been developed, see for instance Ref.\cite{bib:agrawal:pra}. 
 
In this context, Yb-doped fibers are of particular interest for many applications since their quantum defect is small, which contributes to reducing thermal effects, and allows for developing high-power amplifiers and lasers, especially if large-mode-area or multimode (MM) fibers are used. A thorough discussion of Yb-doped fiber amplifiers can be found in Ref.\cite{bib:lindberg:scirep}, whereas static and dynamic instabilities due to thermooptic effects are analysed in Ref.\cite{bib:Laegsgaard:19}.
 
An interesting aspect is that, in a MM fiber amplifier, the beating among modes can create a thermally induced refractive index grating, which in turn leads to modal energy transfer and causes the well known transverse mode instabilities \cite{bib:stihler:lsa}. The amplification of coherent signals in MM fiber amplifiers cannot be merely limited to the study of their overall power throughput; it is essential to consider transversally resolved rate equations, since the light intensity pattern can lead to different amplification regimes, and it has a substantial impact on the quality of the output beam \cite{bib:Jauregui:11}.

Remarkably, steering of the output beam from a MM active fiber could be obtained by properly shaping the input beam phase-front via a spatial light modulator (such as a segmented deformable mirror). This was achieved by measuring the relevant transmission matrix, and by implementing appropriate wavefront control algorithms \cite{bib:Florentin:19,bib:Florentin:lsa}.

On the other hand, in recent years there has been a resurgence of research interest in nonlinear optical effects in multimode fibers. Among these, spatial beam self-cleaning (BSC) due to intermodal four-wave mixing via the Kerr effect was first observed in passive graded-index (GRIN) fibers \cite{bib:Krupa:17}. Soon thereafter, BSC was also reported in an Yb-doped fiber amplifier with a quasi step-index profile \cite{bib:Guenard:17}. Although the radial evolution of the refractive index profile was not following a parabolic profile, it was possible to observe BSC at $1064\,\text{nm}$, both in the presence or in the absence of an additional forward diode pump at $940\,\text{nm}$, leading to signal gain of about $13\,\text{dB}$ \cite{bib:Guenard:17}, or to signal loss of $7.3\,\text{dB}$ owing to Yb ground state absorption, respectively. 
This experimental work was also extended, for a quasi-step index MM fiber, to the case of a coupled-cavity composed MM laser \cite{bib:Guenard:17b} and, more recently, to Yb-doped fiber tapers with graded-index profiles \cite{bib:Niang:19,bib:Alioune:taperyb}.

The numerical simulation of complex beam shaping combining MM propagation, gain/loss mechanisms and the Kerr effect has recently been discussed in the context of spatio-temporal mode-locked fiber lasers \cite{bib:Wright:Science}. This has permitted to point 
out the role played by gain saturation as a modal coupling mechanism for locking both longitudinal and transverse laser modes (see supplementary material of Ref.\cite{bib:Wright:Science}).
In addition, the generation of self-similar pulses in all-normal dispersion mode-locked fiber lasers has been experimentally and numerically studied in Ref.\cite{bib:Tegin:19}.

In this work we introduce a numerical model, suitable for describing spatial beam cleaning effects in active GRIN MM fibers. Our simulation results fully support earlier experimental findings using Yb-doped MM fibers, showing that dissipative MM fibers are a convenient and important platform for obtaining BSC. Moreover, our model permits to design and optimize novel active fiber structures and amplification schemes based on BSC; for example, by introducing a suitable transverse gain/loss distribution in the fiber core.
Our manuscript is organized as follows: Section 2 explains the mathematical model for gain in Yb-doped MM fibers; Section 3 and 4 describe the dynamics in Yb-doped MM fibers, both in the absence of any pumping, and with a forward pump, respectively; Section 5 extends the analysis of BSC and self-imaging to Yb-doped MM tapers; finally, Section 6 draws our conclusions.

\section{Distributed gain model}

Our goal is to numerically study signal beam evolution in Yb-doped MM fibers,
under the combined action of the Kerr effect and gain due to Yb ions, with continuous wave (CW) diode pumping.
For efficient pumping, we consider a double cladding active fiber structure: owing to the presence of thousands of modes in the inner cladding, the pump distribution is uniform in the transverse dimension.

Previous experiments with either passive \cite{bib:Krupa:17} or active MM fibers \cite{bib:Guenard:17} have taught us that, when using signal powers higher than a few tens of kW, a reshaping of the transverse beam profile is obtained after 2-3 m of propagation. As a matter of fact, as the signal power grows larger, one observes a progressive evolution of the output speckled pattern, which is typical of the MM regime, towards a single bell-shaped profile. For such a reason, our numerical model must necessarily include, besides the dependence of the field on the transverse coordinates, a longitudinal fiber perturbation which breaks the radial symmetry of the beam and leads, in the linear regime, to a speckled output pattern as a result of MM interference \cite{bib:PhysRevLett.122.123902,bib:pra:garnier:2019,bib:prl:baudin:2020}. 
We underline that the signal evolution cannot be only modeled in terms of mode powers, since keeping track of the phase evolution is necessary for properly modelling linear and nonlinear mode coupling.

The model presented herein combines transversally resolved rate equations, including the pump power $P_P(z)$, with a nonlinear beam propagation equation for the signal electric field $E(x,y,z)$.
When considering forward pumping, the pump evolution is described by the equation
\begin{equation}
\begin{split}
\frac{dP_P}{dz}=&\left[
(\sigma_{AP}+\sigma_{EP})\frac{\int_S N_2(x,y,z)dS}{A_{Cl}} \right. \\
& \left. -\sigma_{AP}\frac{\int_S N_T(x,y)dS}{A_{Cl}}
\right]
P_P-\alpha_P P_P
\end{split}
\label{eq:proppump}
\end{equation}
where $\sigma_{AP}$ and $\sigma_{EP}$ are absorption and emission cross-sections at the pump wavelength, $N_1$ ($N_2$) is the ground (excited) Yb ion density, $N_T=N_1+N_2$ is the total Yb density, $A_{Cl}$ is the inner cladding area, and $\alpha_{P}$ is the linear pump absorption coefficient. Since we consider a uniformly distributed pump (i.e., it does not depend on $x,y$) in the inner guiding cladding, the pump intensity reads as $I_P(z)=P_P(z)/A_{Cl}$. In the previous equation the local densities of ions in the ground and excited states depend on the three spatial coordinates since, as we will see, it is important to take into account the fact that $N_1$ and $N_2$ vary not only along the longitudinal coordinate $z$, but also in the transverse $x,y$ plane. As a result, the two integrals for the ion densities are calculated on the cross-section $S$, where doping is present. In the present work, we assume that doping is only present in the core, where it is uniformly distributed. Therefore, $N_T$ is constant, and it vanishes outside the core. 

The pump is CW, and we consider the steady-state amplification regime. Nevertheless, when writing the rate equations, we must take into account that the signal pulses have a duration $T_{S}$, a repetition rate $1/T_{R}$ (where $T_{R}$ is the separation between two consecutive pulses) and a time-dependent intensity $I_S(x,y,z,t)$.
We can define the time-averaged signal intensity as $\bar{I}_S(x,y,z)=1/T_{R}\int_{T_{R}} I(x,y,z,t)\,dt\simeq \max_t[I(x,y,z,t)]\,T_{S}/T_{R}$: this is an approximation in this context, although it is often adopted in fiber lasers
\cite{bib:runge:2014}.
It follows that the density of excited ions $N_2(x,y,z)$ can be explicitly written as a function of the pump and the (time-averaged) signal intensities \cite{bib:Laegsgaard:19}:
\begin{equation}
N_2(x,y,z)=\frac{N_T \left(\sigma_{AP}I_P(z)+\sigma_{AS}\bar{I}_S(x,y,z)\lambda_S/\lambda_P\right)}
{\sigma_{AEP}I_P(z)+\sigma_{AES}\bar{I}_S(x,y,z)\lambda_S/\lambda_P+P_\tau}
\label{eq:N2distr}
\end{equation}
where $\sigma_{AEP}=\sigma_{AP}+\sigma_{EP}$, $\sigma_{AES}=\sigma_{AS}+\sigma_{ES}$, $\tau$ is the excited level lifetime, $P_\tau=hc/(\lambda_P\tau)$, $h$ is the Planck's constant and $c$ is the speed of light.

The complex envelope of the signal electric field  $E(x,y,z)$ is normalized in such a way that its intensity is $I(x,y,z)=|E(x,y,z)|^2$; the field evolution is governed by the following equation \cite{bib:Laegsgaard:19}:
\begin{equation}
\begin{split}
\frac{\partial E}{\partial z}=&\frac{i}{2k_0}\bigg\{\nabla^2_\perp E
+\Big[k^2-k_0^2\Big]E\bigg\}
+i\gamma|E|^2E
-\frac{\alpha_S}{2}E \\
&+\frac{1}{2}\bigg[\Big(\sigma_{AS}+\sigma_{ES}\Big)N_2-\sigma_{AS}N_T\bigg]E
\end{split}
\label{eq:propgain}
\end{equation}
where $k_0=2\pi n_0/\lambda_S$ is the reference propagation constant, $k=2\pi\, n(x,y)/\lambda_s $, and $n(x,y)$ is the refractive index distribution. For an ideal GRIN fiber, $n(x,y)$ is well approximated by a circular paraboloid $n^2=n_0^2\left(1-2\Delta \frac{x^2+y^2}{R^2}\right)$, where the reference index is the maximum core refractive index $n_{0}=n_{Co}$, $n_{Cl}$ is the  cladding index, and $\Delta=(n_{Co}^2-n_{Cl}^2)/2 n_{Co}^2$. 
In Eq.(\ref{eq:propgain}), the term in braces describes diffraction and the guiding structure, the term $i\gamma|E|^2E$ represents the instantaneous Kerr nonlinearity, $\alpha_S$ is the linear loss coefficient of the silica glass host, and the last term accounts for the gain $g(x,y,z)$ introduced by the population inversion, which reads as
\begin{equation}
g(x,y,z)=\Big(\sigma_{AS}+\sigma_{ES}\Big)N_2(x,y,z)-\sigma_{AS}N_T(x,y,z)
\label{eq:fieldgain}
\end{equation}
It is worth observing that even if the Kerr coefficient and the doping are homogeneous over the fiber cross-section, the intensity dependent refractive index and the gain function depend on the transverse coordinates, and can lead to an exchange of power among modes carried by the MM fiber.

Moreover, the refractive index distribution can be perturbed by the population inversion of the Yb ions. The variation of the refractive index $\delta n$ depends linearly on the population of the excited state, as described by the equation
\begin{equation}
\delta n=\frac{2\pi}{n_0}\left(\frac{n_0^2+2}{3}\right)^2 \Delta p \,N_2    
\label{eq:deltansat}
\end{equation}
where $\Delta p=1.2\times 10^{-26}$ cm$^3$ is the polarizability difference of the Yb$^{3+}$ ions in the excited and ground states, respectively \cite{bib:Fotiadi:2010}. We also assume that the Yb-doped glass, for the selected pump and signal wavelengths, behaves as a quasi-three-level system \cite{bib:Pask:95,bib:Paschotta:1997}, and that the maximum fraction of inverted population is $N_2/N_T=0.5$ (since $\sigma_{AP}=\sigma_{EP}$): by considering $N_T=9\times 10^{19} \,\text{cm}^{-3}$ and a maximally inverted population $N_2=4.5\times 10^{19} \,\text{cm}^{-3}$, the resulting variation is $\delta n=4.44\times 10^{-6}$. In other words, $\delta n=8.88\times 10^{-6}N_2/N_T$, where $N_2/N_T$ is the fractional inverted population.
In GRIN fibers the self-imaging effect is responsible for the formation of a Kerr-induced index grating \cite{bib:Longhi:03,bib:Agrawal:19}, which also leads to a periodic modulation of $N_2$ according to Eq.(\ref{eq:N2distr}), hence of $\delta n$. 

In order to investigate how this refractive index modulation eventually interferes with the Kerr-induced index grating, in Eq. (\ref{eq:propgain}) we may replace $n(x,y)$ by $n(x,y)+\delta n(x,y,z)$: this part of the study is addressed at the end of section \ref{sec:mat}.
It is also worth observing that a local change in the fiber temperature $\delta T$ of just a few Kelvin can lead to an increase in the refractive index which is larger than that expected from population inversion, since $\delta n \simeq 1.2\times 10^{-5} \delta T$; however, this would lead to refractive index fluctuations on a longitudinal scale which is longer than the self-imaging period. We may thus neglect this contribution in the current numerical analysis, and we plan to address this issue in subsequent studies.

It is also customary to express the guided signal field by using the base provided by the set of Hermite-Gauss guided modes $\psi_{h,k}(x,y)$, which are the eigensolutions of the linear problem 
$\nabla^2_\perp E
+\left[k^2-k_0^2\right]E=0$ for the GRIN fiber \cite{bib:Ghatak:98}.

In the next sections we develop our simulations for an ideal sample case, whose set of parameter values are summarized for simplicity in Table \ref{tab1}.
Whenever necessary, in some simulations the parameters will be modified (for instance, we may neglect the Kerr effect), as explained in the corresponding sections. 

\begin{table}[h!]
  \begin{tabular}{|l|l|l|}
  \hline
  Symbol & Parameter & Value \\
  \hline
  $\lambda_{P}$ & Pump wavelength & $979$ nm\\
  $\lambda_{S}$ & Signal wavelength & $1064$ nm\\
  $T_{S}$ & Signal pulse duration & $500$ ps\\
  $D$ & Signal beam diameter FWHMI & $40$ $\mu$m\\
  $\sigma_{AP}$ & Pump absorption cross section & $2.358\times10^{-24}$ m$^2$\\
  $\sigma_{EP}$ & Pump emission cross section & $2.358\times10^{-24}$ m$^2$\\
  $\sigma_{AS}$ & Signal absorption cross section & $5\times10^{-27}$ m$^2$\\
  $\sigma_{ES}$ & Signal emission cross section & $2.89\times10^{-25}$ m$^2$\\
  $\alpha_{P}$ & Linear absorption (pump) & $1.1$ dB/km\\
  $\alpha_{S}$ & Linear absorption (signal) & $1.1$ dB/km\\
  $\tau$ & Radiative lifetime & $1$ ms\\
  $N_{T}$ & Yb$^{3+}$ concentration & $9.0\times10^{25}$ m$^{-3}$\\
  $n_{Co}$ & Maximum core refractive index & $1.47$\\
  $n_{Cl}$ & Cladding refractive index & $1.457$\\
  $R$ & Core radius & $26$ $\mu$m\\
  $A_{Cl}$ & Inner cladding area & $7.06\times10^{-8}$ m$^2$\\
  $n_2$ & Kerr nonlinear coefficient & $2.6 \times10^{-20}$ m$^2/W$\\
  
  \hline
  \end{tabular}
\caption{Parameters used in modelling and numerical simulations}
\label{tab1}
\end{table}

The disorder along the propagation, which is responsible for the formation of a speckled output intensity pattern in the linear regime, and which may also lead to an acceleration of the beam cleaning process in the nonlinear regime \cite{bib:PhysRevLett.122.123902}, is implemented in the GRIN fiber with a core radius $R=26$ $\mu$m and $\Delta=8.8\times10^{-3}$ by considering randomly oriented elliptical deformations of the fiber core, with a maximum excursion of the axes of 0.15 $\mu$m, and a coarse 5 mm pitch for the ellipse rotation.

\section{Linear and nonlinear dynamics in Yb$^{3+}$ doped fiber in the absence of pumping}

First, by means of an example, we numerically confirm that, whenever the input beam peak intensity grows larger than a certain threshold value, spatial beam self-cleaning can indeed be obtained in an undoped, passive GRIN fiber.
\begin{figure}[h!]
\centering\includegraphics[width=\linewidth]{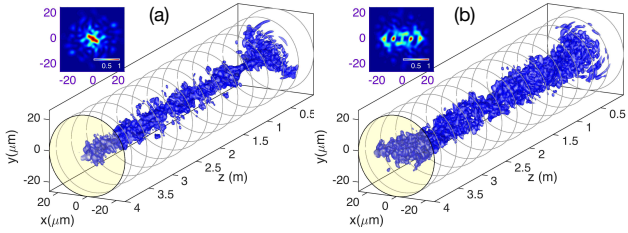}
\caption{(a) Beam cleaning in a passive GRIN fiber with input signal peak intensity $5\,\text{GW/cm}^2$; (b) propagation with input signal peak intensity $0.5\,\text{GW/cm}^2$.}
\label{fig:nodoped}
\end{figure}
Let us consider the parameters of Table \ref{tab1} and an input Gaussian signal beam with a peak intensity in the time domain of 5 GW/cm$^2$. The corresponding numerically calculated evolution (see the iso-intensity surface at half maximum intensity in Fig.\ref{fig:nodoped}(a)) shows that a transition, from a wide and speckled intensity pattern into a cleaned beam, occurs after about $0.5\,\text{m}$. With ten times less input intensity (see Fig.\ref{fig:nodoped}(b)), this beam cleaning effect substantially vanishes: the transverse pattern remains speckled over the full propagation distance of $4\,\text{m}$, as it occurs when the simulation is carried out in the absence of the Kerr effect (i.e., we set $n_2=0$).

\begin{figure}[h!]
\centering\includegraphics[width=\linewidth]{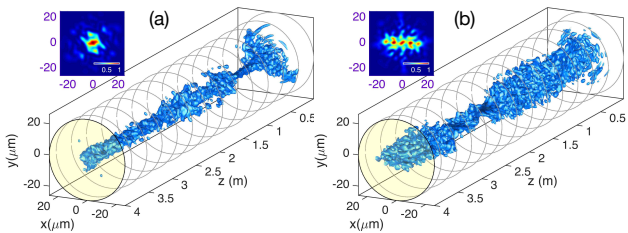}
\caption{(a) Beam cleaning in a nonlinear Yb-doped GRIN fiber, in absence of pump for an input signal maximum intensity $5\,\text{GW/cm}^2$; (b) same propagation after turning off the Kerr effect ($n_2=0$).}
\label{fig:doped_nopump_intensity}
\end{figure}

\begin{figure}[h!]
\centering\includegraphics[width=\linewidth]{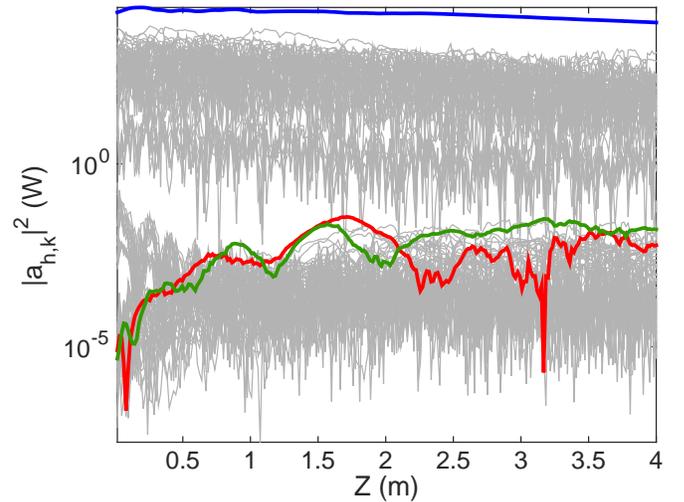}
\caption{Beam cleaning in a nonlinear Yb-doped GRIN fiber, in the absence of pump: evolution of the modal power $|a_{h,k}|^2$ ( $h,k \in [0,9] $) in a logarithmic scale for an input signal peak intensity $5\,\text{GW/cm}^2$ (blue line refers to $a_{0,0}$, red to $a_{0,1}$, green to $a_{1,0}$).}
\label{fig:doped_nopump_modes}
\end{figure}

When one includes a uniform Yb doping (while keeping the same graded index profile), and in the absence of a pump, i.e., with $P_P(0)=0$, the GRIN fiber exhibits high losses, owing to ground state absorption by Yb ions.
Nevertheless, quite surprising Figure \ref{fig:doped_nopump_intensity} shows that, in spite of the additional loss, it is still possible to observe beam self-cleaning in a doped un-pumped fiber, with the same input signal peak intensity of $5$ GW/cm$^2$ (see panel (a)) as in Fig.\ref{fig:nodoped}(a). Obviously, beam cleaning vanishes whenever the simulation does not include the Kerr effect, i.e., with $n_2=0$ (panel (b)). These numerical simulations are in qualitative agreement with experimental results of Ref.\cite{bib:Guenard:17}, that were carried out for a slightly different fiber geometry.
Figure \ref{fig:doped_nopump_modes} reveals the evolution with distance of the beam modal composition, corresponding to the case of Fig.\ref{fig:doped_nopump_intensity}(a), where $a_{h,k}$ is the amplitude of the mode $\psi_{h,k}$. Although the signal drops at a rate of $1.9\,\text{dB/m}$, owing to linear losses due to the absence of a pump for inverting the population of the active medium, we still observe that a significant nonlinear energy exchange among the modes occurs over the first few meters of propagation.
In particular, the blue curve in Fig.\ref{fig:doped_nopump_intensity}(a) shows the power carried by the fundamental mode $\psi_{0,0}$, whereas the red and green curves indicate the power carried by modes $\psi_{0,1}$ and $\psi_{1,0}$, respectively: these two modes are poorly excited, because of the symmetry of the input condition. The other modes $\psi_{h,k}$ are plotted in grey (we considered the projections with $h,k\in[0,9]$).

\begin{figure}[h!]
\centering\includegraphics[width=\linewidth]{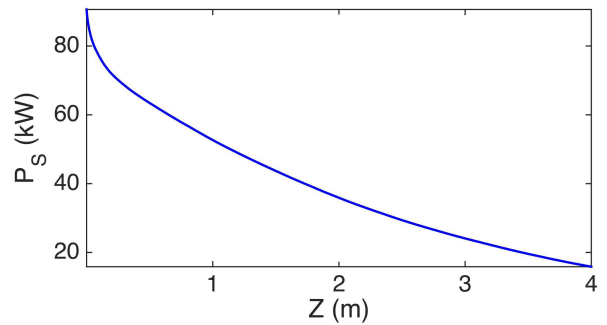}
\caption{Signal power along the Yb-doped GRIN fiber in the absence of pump.}
\label{fig:doped_Powers_nopump}
\end{figure}

It is also instructive to calculate the corresponding signal power evolution, as displayed in Fig.\ref{fig:doped_Powers_nopump}, where $P_S(z)=\int_S|E(x,y,z)|^2 dS$. It is worth underlining that nonlinear effects occur substantially before the signal is attenuated along its propagation.

\section{Linear and nonlinear dynamics in doped fiber with forward pumping}

\begin{figure}[h!]
\centering\includegraphics[width=\linewidth]{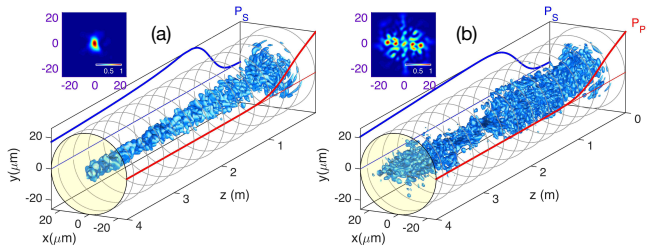}
\caption{(a) Beam cleaning in a nonlinear Yb-doped GRIN fiber: forward pump power of 20 W, and input signal peak intensity $0.5\,\text{GW/cm}^2$; (b) propagation when turning off the Kerr effect ($n_2=0$).}
\label{fig:doped_intensity}
\end{figure}

In this Section, we numerically simulate the dynamics of the signal beam amplification when the pump is switched on.
The corresponding iso-intensity levels are shown in Fig.\ref{fig:doped_intensity}: as it can be seen, for a pump power level of 20 W, beam cleaning is obtained with a signal of 0.5 GW/cm$^2$ peak intensity only. Note that with such input signal level it was not possible to obtain beam cleaning in a passive GRIN fiber (see panel (b) of Fig.\ref{fig:nodoped}).
We may thus numerically confirm the behavior reported in the experiments of Ref.\cite{bib:Guenard:17}: specifically, a reduction of the input threshold power for beam cleaning in an active fiber, owing to signal amplification which accompanies nonlinear beam reshaping.

\begin{figure}[h!]
\centering\includegraphics[width=\linewidth]{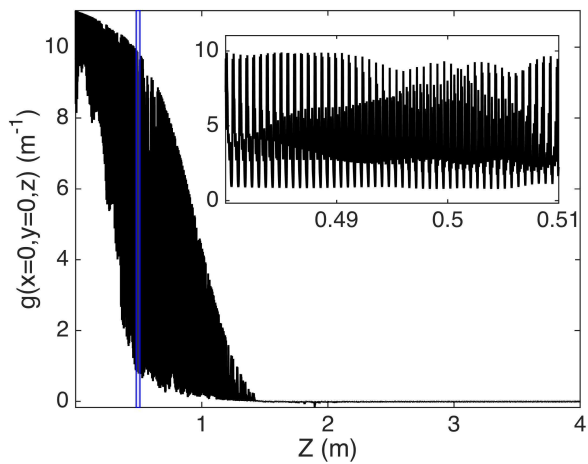}
\caption{Signal gain evolution in a nonlinear Yb-doped GRIN fiber: forward pump power of 20 W and input signal peak intensity 0.5 GW/cm$^2$. The inset is a zoom of the region highlighted by the blue vertical lines.}
\label{fig:doped_gain}
\end{figure}

The field gain $g/2$ (as defined in Eq. (\ref{eq:fieldgain})) experienced by the complex electric field $E$ is neither uniform along the fiber (due to the progressive pump absorption), nor uniform in the transverse domain, owing to the different population inversion across the beam cross-section. 
In particular, gain saturation is most pronounced at the point of minimum waist for the signal beam. The well-known periodic evolution of the beam in a parabolic refractive index profile causes then a periodic fluctuation of the signal intensity. In turn, this gives rise to a periodic evolution of the inverted fraction $N_2(x,y,z)$ along the coordinate $z$. To exemplify this effect, in Fig.\ref{fig:doped_gain} we show the evolution of the gain along the fiber axis $g(x=0,y=0,z)$. The inset is a detail of the evolution over a shorter fiber span, as highlighted by the blue bar.
Maximum gain is experienced at the beginning of the propagation, and then gain gradually drops as the pump is absorbed during its propagation. However, the most striking feature of the $g$ evolution is the fast varying oscillation on a sub-millimetric scale, which is due to the coherent beating of all modes with the self-imaging period of $\Lambda=\pi R / \sqrt{2 \Delta}$ (with the present parameters, $\Lambda=0.6\,\text{mm}$). 

\begin{figure}[h!]
\centering\includegraphics[width=\linewidth]{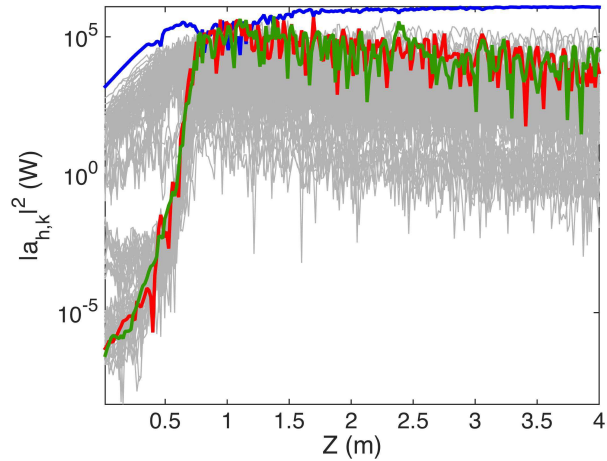}
\caption{Beam cleaning in a nonlinear Yb-doped GRIN fiber: mode power evolution on a logarithmic scale. Forward pump of $20\,\text{W}$ power, and input signal peak intensity $0.5\,\text{GW/cm}^2$ (the blue line refers to $a_{0,0}$, the red to $a_{0,1}$, and the green to $a_{1,0}$).}
\label{fig:doped_modes}
\end{figure}

A better understanding of the signal dynamics can be gained by observing the evolution of the power carried by the different modes $|a_{h,k}|^2$ (see Fig.\ref{fig:doped_modes}). Specifically, it is possible to notice that, although all modes are amplified, their gain is not uniformly distributed among them. This is surprising, since in the absence of nonlinearity and gain saturation, the uniform doping distribution coupled with the cladding pumping scheme would lead to the same gain per mode.
This process culminates in a strong nonlinear reshaping nearby $z=1\,\text{m}$.

Along the propagation of the signal beam, amplification increases nonlinear mode mixing, and reduces the nonlinear length. As a result, in Fig.\ref{fig:doped_modes} we observe a rapid growth of both low and high-order modes (see for instance the red and green curves, referring to modes $a_{1,0}$,$a_{0,1}$). Note that these very modes are poorly excited in the initial steps of the propagation. In parallel, nearby $z=1\,\text{m}$ in Fig.\ref{fig:doped_modes}, we notice a local decrease of the fundamental mode power (blue curve), before a new amplification stage, which can be ascribed to power flow from high-order modes. Thus, high-order modes can efficiently absorb the gain distributed on the outer edge of the fiber core, before  transferring it to the fundamental mode by four-wave mixing. For $z>1\,\text{m}$, the power of the fundamental mode grows larger. Such multi-step nonlinear amplification mechanism, involving  the amplification of many modes coupled with nonlinear power transfer toward the fundamental mode via the Kerr effect, provides an original light amplification scheme for nonlinear multimode fiber amplifiers. 

\begin{figure}[h!]
\centering\includegraphics[width=\linewidth]{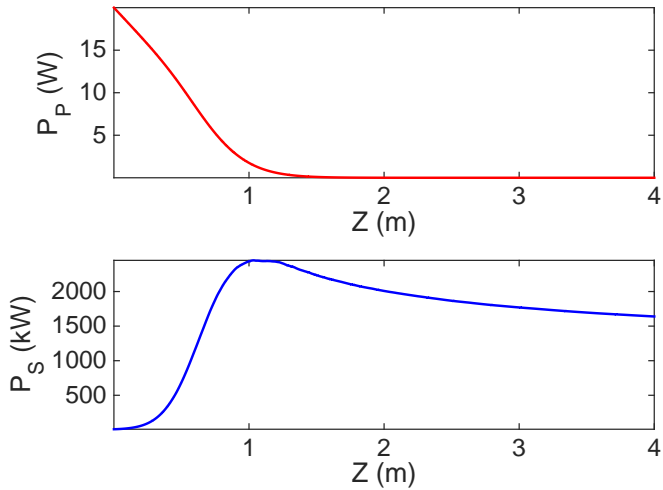}
\caption{Powers of pump and signal along the fiber: $\bar{I}_S(0,0,0)=0.5$ GW/cm$^2$ and input pump power $P_P(0)=20\,\text{W}$.}
\label{fig:doped_Powers}
\end{figure}

The situation can be possibly clarified by inspecting the evolution of pump and signal powers, as reported in Fig.\ref{fig:doped_Powers}. The strong nonlinear reshaping at $z=1\,\text{m}$ 
corresponds to the maximum amplification of the signal, whose peak power reaches above $2.5\,\text{MW}$. Moreover, at that distance the pump is entirely absorbed, so that over the remaining propagation distance the signal power is reduced. We can conclude that the Kerr effect may provide an efficient mechanism of nonlinear mode scrambling (especially close to the threshold for beam collapse) and power transfer. We also observed that there is a well-defined optimal amplifier length for obtaining a maximum amplification of the signal.

\section{Multimode Active Tapers}
\label{sec:mat}

An important category of fiber amplifiers is provided by fiber tapers, where both cladding and core radius $R(z)$ vary along the amplifier length. In this Section, we support by numerical simulations the results of a series of experiments on nonlinear spectral broadening and beam reshaping in both passive and active MM fiber tapers \cite{Eftekhar2019,bib:Krupa:19,bib:Hansson:20,bib:paperMarioPRAPP}. 
Here we limit our analysis to the spatial coordinates: in this case, Eq. (\ref{eq:propgain}) can be easily extended to tapers, where the radius $R(z)$ varies from a value $R_{IN}$ to a value $R_{OUT}$. For the ease of discussion, we model the evolution of the core radius by means of a linear function $R(z)=R_{OUT}+(z-L)(R_{OUT}-R_{IN})/L$, where $L$ is the taper length. In practical implementations of fiber tapers, the variation upon distance of the core radius is not necessarily linear. We refer here to the case $R_{IN}>R_{OUT}$, in order to qualitatively reproduce some of the experimental results of Ref.\cite{bib:Niang:19}. Note that the decrease of core radius with fiber length progressively increases the strength of the fiber nonlinearity, which leads to an acceleration of nonlinear mixing along the beam propagation \cite{Eftekhar2019}.
Similarly to the previous considered cases, here the pump is combined with the signal in the forward direction. Aiming at a comparative analysis with the previous cases, we considered the same set of parameters' values that we used for uniform or $z$-invariant fibers. However, in order to match some of the features of the taper of Ref.\cite{bib:Niang:19}, we have chosen $L=9\,\text{m}$, $R_{IN}=61\,\mu\text{m}$, $R_{OUT}=18\,\mu\text{m}$. Moreover, in our simulation the cladding radius linearly varies from $350\,\mu\text{m}$ down to $90\,\mu\text{m}$. Again, the presence of fiber disorder is introduced by means of local elliptical deformations of the fiber core. Since the core radius changes along the propagation of the signal beam, these deformations are uniformly randomly distributed with a maximal excursion proportional to $R(z)$. In the simulation the maximal deformation on each axis is $5.8\times10^{-3}R(z)$, and with a coarse step of $5\,\text{mm}$.

\begin{figure}[h!]
\centering\includegraphics[width=\linewidth]{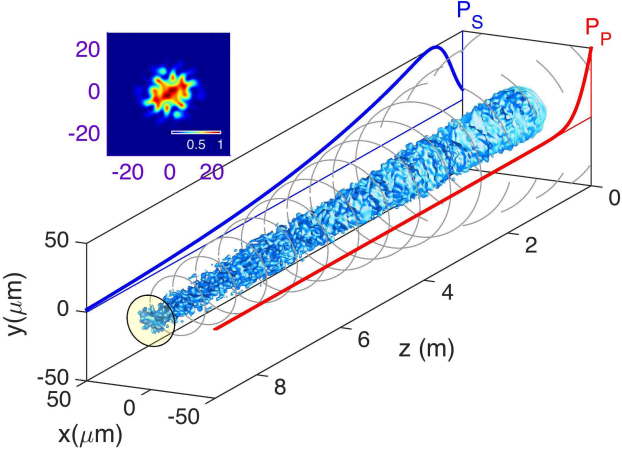}
\caption{Iso-intensity surface at half-maximum intensity in a Yb-doped active fiber taper in the absence of Kerr effect ($n_2=0$). The blue (red) curve reproduces the evolution of the signal (pump) power. The grey circles represent the local size of the fiber core. The inset shows the output beam intensity pattern.}
\label{fig:fiber_taper_nk}
\end{figure}

\begin{figure}[h!]
\centering\includegraphics[width=\linewidth]{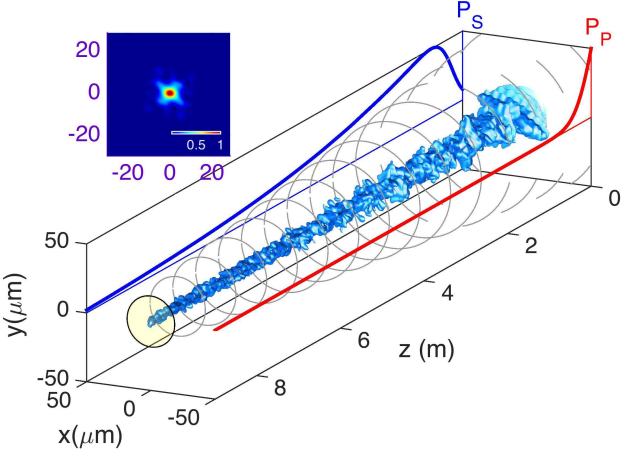}
\caption{Same as in Fig.\ref{fig:fiber_taper_nk}, but in the presence of the Kerr effect.}
\label{fig:fiber_taper}
\end{figure}

The numerical simulation presented in Figure \ref{fig:fiber_taper_nk} illustrates signal beam propagation along the active taper, in the absence of the Kerr effect (i.e., we set $n_2=0$); here the input diode pump power is $20\,\text{W}$, and the signal intensity is $1\,\text{GW/cm}^2$. As can be seen, the output beam exhibits a speckled intensity pattern. As shown in Fig.\ref{fig:fiber_taper}, the result is qualitatively different when adding the contribution of the Kerr effect. Specifically, the simulation shows that the output signal intensity pattern exhibits a large bright spot at its center, surrounded by a weak background.
Note that in the considered case the pump is absorbed over the first meter of fiber, so that for longer distances the signal is experiencing a propagation loss. Nevertheless, Fig.\ref{fig:fiber_taper} clearly illustrates the gradual transition from a speckled to a cleaned beam, when the Kerr effect is switched on.

\begin{figure}[h!]
\centering\includegraphics[width=\linewidth]{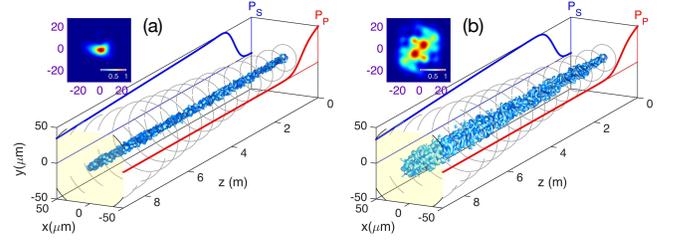}
\caption{Numerical simulation of nonlinear beam propagation in a Yb-doped fiber taper with $R_{IN}=18.5\,\mu \text{m}$ $R_{OUT}=61\,\mu \text{m}$ (the signal propagation direction is inverted with respect to Figs.\ref{fig:fiber_taper_nk},\ref{fig:fiber_taper}); (a) in the presence of the Kerr effect; (b) in the absence of the Kerr effect ($n_2=0$).}
\label{fig:invertedtaper}
\end{figure}

In fiber amplifiers and lasers, fiber tapers are commonly used by propagating the signal in the opposite direction, namely, with $R_{IN}<R_{OUT}$. In fact, if the size of the core increases at the same time that the signal is amplified, the nonlinearity increase due to the growing signal power can be compensated by the decrease of the nonlinear coefficient, because of the larger effective area of the guided beam. An example of the nonlinear dynamics in this configuration for the multimode active taper is illustrated in Fig.\ref{fig:invertedtaper}: here 
$R_{IN}=18\,\mu \text{m}$, $R_{OUT}=61\,\mu \text{m}$, the input beam diameter (FWHMI) is $12.5\,\mu \text{m}$, and the input pump power is $20\,\text{W}$.
In details, panel (a) in Fig.\ref{fig:invertedtaper} illustrates nonlinear propagation in the presence of disorder and Kerr effect. As can be seen, the signal beam width remains substantially unchanged along the taper, in spite of experiencing a drastic increase of the core diameter. For comparison, panel (b) illustrates propagation in the same taper, but in the absence of the Kerr effect (we set $n_2=0$): now, the beam spreads wider along its propagation. Although the present simulations refer to a forward pumping situation, these numerical results substantially agree with the experimental observations of spatial beam self-cleaning in active tapers with decelerating nonlinearity, as reported in Ref.\cite{bib:Alioune:taperyb}, and obtained with a backward pump. 

Since core and cladding radii vary slowly along the taper, a significant change of their values can only be obtained after a distance that is orders of magnitude greater than the typical value of the self-imaging period (which ranges from a few millimeters to a few hundred microns). For this reason, the self-imaging effect can also be observed in short segments of the taper in Fig.\ref{fig:fiber_taper}. Since the period $\Lambda$ is proportional to the core radius $R$ \cite{bib:Agrawal:19}, self-imaging accelerates (i.e., $\Lambda$ decreases) when going from the large core input facet ($\Lambda\sim 1.3\,\text{mm}$) to the small core one ($\Lambda\sim 0.4\,\text{mm}$).

\begin{figure}[h!]
\centering\includegraphics[width=\linewidth]{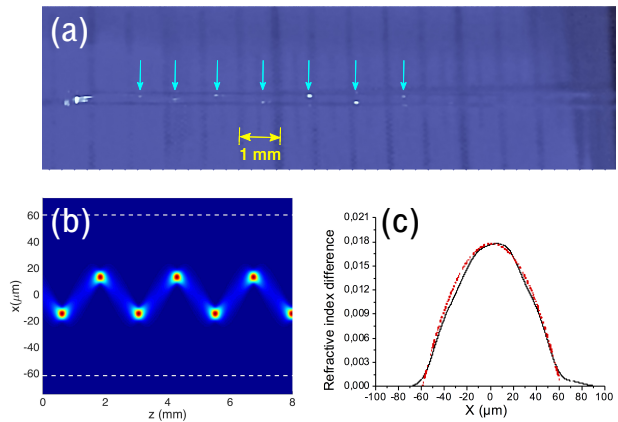}
\caption{(a) Photoluminescence measured by using a pulsed laser at $1030\,\text{nm}$ for a short taper segment with a fiber radius $R=61\,\mu\text{m}$; (b) numerically calculated intensity evolution in a uniform fiber with core radius $R=61\,\mu\text{m}$; (c) measured local refractive index profile.}
\label{fig:SLFIM}
\end{figure}

As an example, we may consider the Yb-doped fiber that was used in Ref.\cite{bib:Niang:19}. Panel (a) of Fig.\ref{fig:SLFIM} shows the experimental photoluminescence, that is observed at points where the maxima of intensity are reached: here the spacing between two consecutive points is $1.2\,\text{mm}$. These results are obtained by using an experimental setup which is similar to that described in Refs.\cite{bib:Krupa:19,bib:Hansson:20,bib:paperMarioPRAPP,bib:paperMarioMULTIPHOTON}. Panel (b) shows the calculated intensity evolution along an uniform fiber segment, i.e., with constant radius $R=61 \,\mu\text{m}$ and a parabolic refractive index profile, corresponding to the experimental profile in panel (c). In such a case, the input Gaussian beam is injected off axis, and the corresponding intensity distribution exhibits a zig-zag trajectory with a period of $2 \Lambda$. As a matter of fact, $\Lambda$ is the self-imaging period for an on-axis symmetric input field only, similarly to what is observed in other multimode waveguides \cite{bib:Bachmann:94,bib:Bachmann:95}.  
With $R=61\,\mu\text{m}$, $\Delta n=0.018$ and assuming for instance $n_{Cl}=1.457$, the calculated self-imaging period is $\Lambda=1.2\,\text{mm}$: this value is close to the experimental result in Fig.\ref{fig:SLFIM}(a), and agrees well with the numerical result in Fig.\ref{fig:SLFIM}(b).

\begin{figure}[h!]
\centering\includegraphics[width=\linewidth]{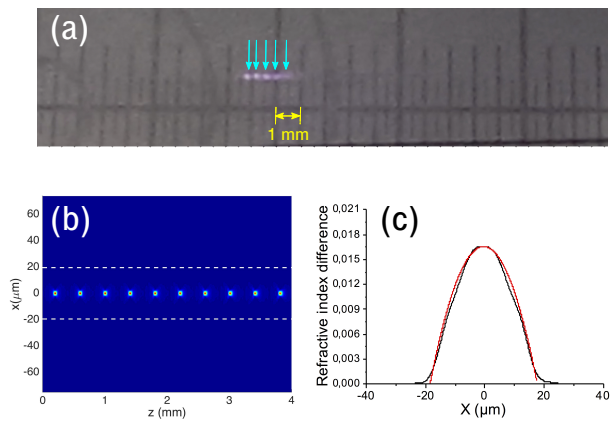}
\caption{(a) Photoluminescence measured by using a pulsed laser at $1030\,\text{nm}$ for a short taper segment with a fiber radius $R=19.5\,\mu\text{m}$; (b) numerically calculated intensity evolution in a uniform fiber having core radius $R=19.5\,\mu\text{m}$; (c) measured local refractive index profile.}
\label{fig:SLFIMsm}
\end{figure}

The signal beam intensity maxima are more closely packed when observing the taper near its small input facet, as illustrated in Fig.\ref{fig:SLFIMsm}: here panel (a) shows the light scattered by photoluminescence. As can be seen, the spacing between two consecutive bright spots reduces down to $0.36\,\text{mm}$, owing to the reduced diameter. Whereas panel (b) illustrates the numerical propagation result, again assuming a segment of constant diameter for a few millimeters of propagation, and panel (c) shows the corresponding measured local refractive index profile. With $R=19.5\,\mu\text{m}$, $\Delta n=0.017$ and $n_{Cl}=1.457$, the expected self-imaging period is $\Lambda=0.4\,\text{mm}$. 
These experimental results agree well with the numerical ones, and with the self-imaging theory, thus extending the analysis presented in Ref.\cite{bib:Krupa:19,bib:Hansson:20,bib:paperMarioPRAPP} from standard 50/125 GRIN MM fibers to the case of Yb-doped MM tapers. 

\begin{figure}[h!]
\centering\includegraphics[width=\linewidth]{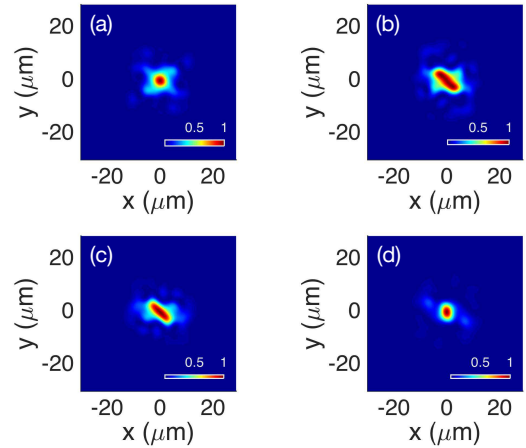}
\caption{Beam shapes at the output of the fiber taper with different strengths of refractive index perturbation $\delta n$, caused by the population inversion of Yb ions. (a) $\delta n=1\times 10^{-6}N_2/N_T$; (b) $\delta n=4\times 10^{-6}N_2/N_T$; (c) $\delta n=6\times 10^{-6}N_2/N_T$; (d) $\delta n=9\times 10^{-6}N_2/N_T$. The remaining parameters are those of Fig.\ref{fig:fiber_taper}.}
\label{fig:satdn}
\end{figure}

At this point we may briefly address the relevant problem of the local refractive index perturbations that are induced by the population inversion in the active fiber medium.
Figure \ref{fig:satdn} analyses the additional impact of the refractive index perturbation $\delta n$, which is caused by the population inversion of Yb ions, as described by Eq. (\ref{eq:deltansat}). Here we show the output beam profiles for realistic values of $\delta n$. The simulations have been carried out with the same parameters as in Fig.\ref{fig:fiber_taper}. To simplify our analysis, $\delta n$ is assumed to be proportional to the ratio $N_2/N_T$, with a proprtionality factor ranging from $1\times 10^{-6}$ (panel (a)) up to $9\times 10^{-6}$ (panel (d)). The conclusion of this analysis is that, although population inversion can induce a refractive index perturbation, its impact is not sufficient to reduce the effectiveness of spatial beam cleaning, at least within the considered range of parameters.  

\section{Conclusions}

In this work, we carried out a series of numerical simulations in order to study the onset of nonlinear beam reshaping and cleaning in MM GRIN fiber amplifiers. At variance with single-mode fibers, GRIN fibers lead to a periodic evolution of the signal intensity, which, in turn, periodically modulates the population inversion. The beam cleaning effects observed in our numerical simulations agree well with experimentally reported spatial beam self-cleaning in Refs.\cite{bib:Guenard:17,bib:Guenard:17b,bib:Niang:19,bib:Alioune:taperyb}, and add a new building block to the study of this interesting nonlinear phenomenon. In this work, as well as in the reported experiments, we have shown how BSC can be observed even in dissipative fiber systems, that is when the signal power evolves (under the effect of either linear absorption or amplification), that is in absence of conservation of the signal power. Peculiar properties of the associated nonlinear beam evolution have been pointed out, specifically in connection with the contribution of the active medium, and its non-uniform transverse population distribution, whose detailed analysis deserves further studies.  

\section*{Funding}
H2020 European Research Council (ERC) (grant
No. 740355); Arianegroup (X-LAS); Polish National Agency for Academic Exchange (POLONIUM, grant No. BPN/BFR/2021/1/00013) and French-Polish Partnership Hubert Curien (grant No. 48161TH).

\section*{Acknowledgments}
We acknowledge useful discussions and suggestions by Dr. Alain Barth\'el\'emy. 

\section*{Disclosures}
The authors declare no conflicts of interest. Data underlying the results presented in this paper are not publicly available at this time but may be obtained from the authors upon reasonable request.


\bibliography{referencesarxiv.bib}

\end{document}